\documentclass[conference]{IEEEtran}
\IEEEoverridecommandlockouts
\usepackage{cite}
\usepackage{amsmath,amssymb,amsfonts}
\usepackage{algorithmic}
\usepackage{graphicx}
\usepackage{textcomp}
\usepackage{xcolor}
\def\BibTeX{{\rm B\kern-.05em{\sc i\kern-.025em b}\kern-.08em
    T\kern-.1667em\lower.7ex\hbox{E}\kern-.125emX}}

\usepackage{tw}

\begin{document}

\title{Probabilistic Shaping for Asymmetric Channels and Low-Density Parity-Check Codes
\thanks{This work was supported by the German Research Foundation (DFG) through projects 390777439 and 509917421 and by the National Science Foundation (NSF) grant 1911166. Any opinions, findings, and conclusions or recommendations expressed in this material are those of the authors and do not necessarily reflect the views of the NSF.}
}

\author{\IEEEauthorblockN{Thomas Wiegart\IEEEauthorrefmark{1}, Linfang Wang\IEEEauthorrefmark{2}, Diego Lentner\IEEEauthorrefmark{1}, Richard D. Wesel\IEEEauthorrefmark{2}}
\IEEEauthorblockA{\IEEEauthorrefmark{1}\textit{Institute for Communications Engineering}, Technical University of Munich, Germany \\
\IEEEauthorrefmark{2}\textit{Communications Systems Laboratory}, University of California Los Angeles, USA \\
emails: thomas.wiegart@tum.de, lfwang@g.ucla.edu, diego.lentner@tum.de, wesel@ucla.edu}
}

\maketitle

\begin{abstract}
An algorithm is proposed to encode low-density parity-check (LDPC) codes into codewords with a non-uniform distribution. This enables power-efficient signalling for asymmetric channels. We show gains of $\SI{0.9}{dB}$ for additive white Gaussian noise (AWGN) channels with on-off keying modulation using 5G LDPC codes.
\end{abstract}

\begin{IEEEkeywords}
LDPC codes, probabilistic shaping, forward-error correction, asymmetric signalling.
\end{IEEEkeywords}

\section{Introduction}

This paper explores \ac{LDPC} coded communications with high spectral efficiency over asymmetric channels such as the \ac{AWGN} channel with \ac{OOK}. To approach capacity, one must usually use shaping so that signal points are not equally likely (probabilistic shaping), not uniformly spaced (geometric shaping), or both \cite{gallager1968information,forney1992trellis,Kschischangoptimaldist,laroia1994optimal,xiao2021finite}.

A popular technique called \ac{PAS} \cite{boecherer_bandwidth15,pas_bocherer_2} uses a \ac{DM} to perform shaping before the \ac{FEC} encoder, e.g., a \ac{CCDM} \cite{CCDM_schulte}. PAS factors the target channel-input distribution as $P_X = P_S \cdot P_A$ where $P_S$ is the uniform distribution used for parity bits and $P_A$ relates to the input amplitudes. \ac{PAS} can approach capacity for symmetric channels where the capacity-achieving $P_X$ is symmetric. An especially attractive feature of \ac{PAS} is its flexibility: \ac{PAS} can be combined with any (systematic) \ac{FEC} code, it permits fine rate adaptation, and it performs well with long \cite{pas_bocherer_2} and short \cite{Linfang_CRC_TCM_PAS} block lengths.

PAS does not directly apply to channels where the optimal input distribution is asymmetric. For example, for the \ac{AWGN} channel with \ac{OOK} modulation, the authors of \cite{git_protograph19} propose a \ac{TS} scheme that combines FEC with a non-uniform OOK signaling. In the TS scheme, a DM generates non-uniform bits from a uniform message, and a systematic FEC encoder appends uniformly-distributed parity bits.

Another probabilistic shaping scheme for OOK is presented in \cite{wiegart_shaped19} that  uses the method by Honda and Yamamoto \cite{honda_polar13}. Here, polar codes perform joint distribution matching and FEC by using a polar decoder to encode message bits into a subset of the code with the desired distribution. The polar coding scheme (asymptotically) generates OOK symbols with the capacity-achieving distribution and performs better than the TS scheme. The method by Honda and Yamamoto can not be directly extended to other linear block codes. 

There are several schemes to shape LDPC codes by using a decoder to encode; see~\cite{wainwright2009_lowdensity, mondelli_achieve18}. However, these schemes may encode to an invalid codeword under \ac{BP} encoding. The papers \cite{aref2013_approaching, kumar2013_spatially} report that spatial coupling and guided decimation can combat this problem, and the paper \cite{mondelli_achieve18} proposes an outer \ac{FEC} code to correct encoding errors. These workarounds do not guarantee a valid encoding.

The paper \cite{boecherer_probabilistic19} proposes \ac{LLPS} that shapes the codewords of a linear block code with two DMs: The message bits are shaped with a conventional DM and encoded systematically. A \ac{SDM} shapes the parity bits by reserving $\ell$ bits and determining them 
to achieve a non-uniformly distributed codeword. However, the \ac{SDM} requires enumerating all $2^\ell$ combinations of these $\ell$ bits which is infeasible for large $\ell$.

This paper proposes an efficient algorithm to approximate \ac{LLPS} for \ac{LDPC} codes. The algorithm uses systematic encoding and sequentially determines $\ell$ bits of the systematic part of the codeword. It it based on a \ac{BP}-like algorithm on the Tanner graph of the generator matrix. The algorithm is sub-optimal but it has several attractive features: it generates a valid codeword, it can be implemented efficiently, and it provides reasonable power gains.

This paper is organized as follows: Section~\ref{sec:preliminaries} introduces the AWGN channel with OOK modulation, LDPC codes, and \ac{LLPS}. Section~\ref{sec:shapedLDPC} presents the proposed algorithm and Section~\ref{sec:results} provides simulation results. Section~\ref{sec:conclusion} concludes the paper.

\section{Preliminaries} \label{sec:preliminaries}
\subsection{Channel Model} \label{sec:awgn}
Consider the \ac{AWGN} channel with \ac{OOK} modulation:
\begin{equation}
    Y = X + N
\end{equation}
where the transmit symbol $X$ has alphabet $\{0, A\}$ and $N \sim \mathcal{N}(0,\sigma^2)$ is additive Gaussian noise with zero mean and variance $\sigma^2$. We write $p_0 = P_X(0)$ and the \ac{SNR} is
\begin{equation}
    \gamma = \frac{(1-p_0)A^2}{\sigma^2} .
\end{equation}
A non-uniform distribution $P_X$ for $X$ can be beneficial under an average transmit power constraint such as $\operatorname{E}[X^2] \leq P$. We map codeword symbol $0$ to the $0$-symbol and codeword symbol $1$ to the $A$-symbol. %
\subsection{LDPC Codes}
\ac{LDPC} codes \cite{gallager1968information}  are linear block codes with a sparse parity check matrix. They can be represented by a bipartite graph, called a Tanner graph, with \acp{VN} representing the codeword symbols and \acp{CN} representing the parity checks.
\ac{LDPC} codes can be efficiently decoded using \ac{BP} that iteratively exchanges likelihoods of the codeword bits between the check and variable nodes. Let $L_{\text{V}_i\to\text{C}_j}$ be the message from VN $i$ to CN $j$ and $L_{\text{C}_j\to\text{V}_i}$ be the message from CN $j$ to VN $i$. The message update rules at the variable and check nodes are
\begin{align}
    L_{\text{V}_i\to\text{C}_j} &= L_i + \sum_{k \in \cN(\text{V}_i), k \neq j} L_{\text{C}_k\to\text{V}_i} \label{eq:vnupdate}\\ 
    L_{\text{C}_j\to\text{V}_i} &= 2 \tanh^{-1} \left( \prod_{k \in \cN(\text{C}_j), k \neq i} \tanh \left( L_{\text{V}_k\to\text{C}_j}/2 \right) \right) \label{eq:cnupdate}
\end{align}
where $L_i$ is the \ac{LLR} of the code bit associated with VN $i$ and $\cN(\text{V}_i)$ is the index-set of all neighbours of VN $i$, and similarly for $\cN(\text{C}_j)$. %
The \ac{APP} LLR of VN $i$ is
\begin{equation}
    L_i^\text{APP} = L_i + \sum_{k \in \cN(\text{V}_i)} L_{\text{C}_k\to\text{V}_i} \label{eq:LAPP}\\ 
\end{equation}
and used to decide whether code bit $i$ is zero ($L_i^\text{APP}\geq 0$) or one ($L_i^\text{APP}<0$). The \ac{BP} decoder is usually terminated once a valid codeword is found or one reaches a maximum number of iterations.

\subsection{Linear Layered Probabilistic Shaping (LLPS)}

\ac{LLPS} \cite{boecherer_probabilistic19} is an architecture to encode any linear code of length $n_c$ and dimension $k_c$ to non-uniform codewords. The output of a \ac{DM} is encoded systematically. To obtain parity bits with non-uniform distribution with binary entropy $H_2(p_0)$, one needs to determine approximately
\begin{equation}
    \ell \approx (n_c - k_c) \left( \frac{1}{H_2(p_0)} - 1 \right) 
\end{equation}
bits using a \ac{SDM}.

Consider systematic encoding of a vector $\bm{v}$ of $k_c - \ell$ bits output by a \ac{DM}. The parity check matrix $\bm{H}$ is partitioned as $\bm{H} = [\bm{H}_s | \bm{H}_p]$ where $\bm{H}_s$ is a $m \times (k_c - \ell)$ matrix and $\bm{H}_p$ is a full-rank $m \times (m+\ell)$ matrix with $m = n_c - k_c$. Next, calculate the syndrome $\bm{s} = \bm{v} \bm{H}^T_s$ and observe that any $\bm{c} = [\bm{v} | \bm{p}]$ is a valid codeword if $\bm{p}$ is chosen such that $\bm{p} \bm{H}_p^T = \bm{s}$ is fulfilled. There are $2^\ell$ valid solutions for $\bm{p}$ and one can, e.g., choose the one with lowest Hamming weight  $\text{w}_\text{H}(\bm{p})$:
\begin{equation}
    \bm{p} = \underset{\bm{p}' \in \{0,1\}^{m+\ell}}{\operatorname{argmin}} \text{w}_\text{H}(\bm{p}') \quad \text{s.t.} \quad \bm{p}' \bm{H}_p^T = \bm{s} . \label{eq:LLPSopt}
\end{equation}
The authors of \cite{boecherer_probabilistic19} further decompose $\bm{H}_p$ to solve \eqref{eq:LLPSopt} by enumerating over $\ell$ bits. This allows to shape a linear code with a desired distribution but has high complexity since $2^\ell$ possibilities must be enumerated to solve \eqref{eq:LLPSopt}. \ac{LLPS} with \ac{SDM} is thus feasible only for high-rate distribution matching with small $\ell$ (close-to-uniform distributions) and short blocklengths.

\section{Shaped LDPC Codes} \label{sec:shapedLDPC}
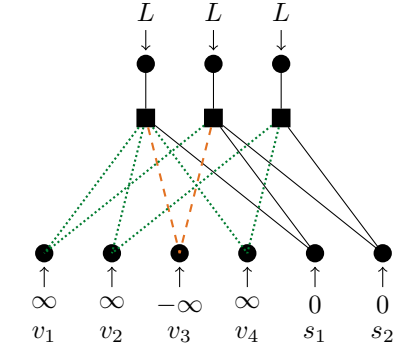
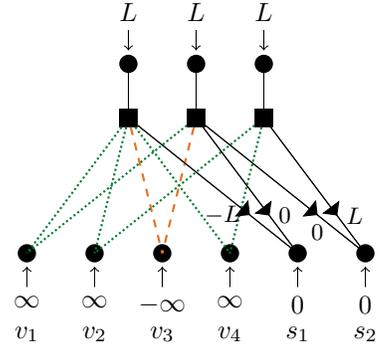
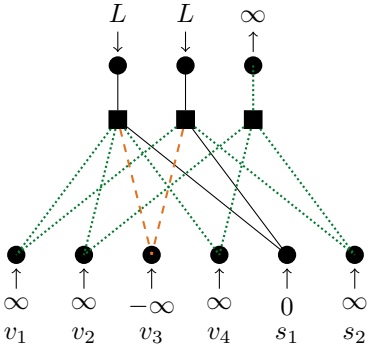
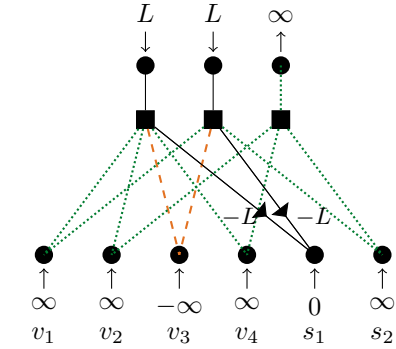
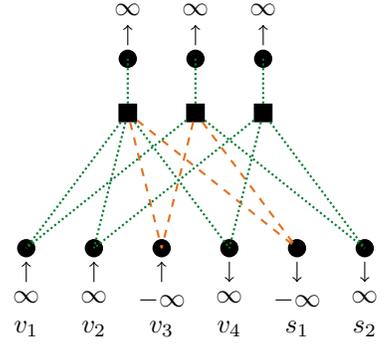
\begin{figure*}[t]
    \centering
    \begin{subfigure}[b]{.3\linewidth}
        \begin{equation*}
				\bm{G} = \begin{bmatrix}
					1 & 0 & 0 & 0 & 0 & 0  & 1 & 1 & 0  \\
					0 & 1 & 0 & 0 & 0 & 0  & 1 & 0 & 1  \\
					0 & 0 & 1 & 0 & 0 & 0  & 1 & 1 & 0  \\
					0 & 0 & 0 & 1 & 0 & 0  & 1 & 0 & 1  \\
					0 & 0 & 0 & 0 & 1 & 0  & 1 & 1 & 0  \\
					0 & 0 & 0 & 0 & 0 & 1  & 0 & 1 & 1  
				\end{bmatrix}
			\end{equation*}
        \caption{Systematic generator matrix of the example code.}
        \label{fig:encgraph:matrix}
    \end{subfigure}
    \hfill
    \begin{subfigure}[b]{.3\linewidth}
        \centering
        \begin{tikzpicture}
\usetikzlibrary{arrows}  

\tikzset{%
          insert new path/.style={%
             insert path={%
                  node[pos=0.7,sloped]{\tikz \draw[#1,thick] (-.2pt,0) -- ++(.2 pt,0);}
                  }
             }
         }
\begin{scope}[scale=0.9, a/.style = {insert new path = {-triangle 90}}]
	\foreach \i in {0,...,5} {
		\fill (\i,0) circle[radius=0.135,fill=black!20] coordinate(vn_\i);
	}
	\draw[<-] ([shift={(0,-0.2)}]vn_0) --++ (0,-0.3) node[below] {$\infty$} node[below=0.4cm] {$v_1$};
	\draw[<-] ([shift={(0,-0.2)}]vn_1) --++ (0,-0.3) node[below] {$\infty$} node[below=0.4cm] {$v_2$};
	\draw[<-] ([shift={(0,-0.2)}]vn_2) --++ (0,-0.3) node[below] {$-\infty$} node[below=0.4cm] {$v_3$};
	\draw[<-] ([shift={(0,-0.2)}]vn_3) --++ (0,-0.3) node[below] {$\infty$} node[below=0.4cm] {$v_4$};
	\draw[<-] ([shift={(0,-0.2)}]vn_4) --++ (0,-0.3) node[below] {$0$} node[below=0.4cm] {$s_1$};
	\draw[<-] ([shift={(0,-0.2)}]vn_5) --++ (0,-0.3) node[below] {$0$} node[below=0.4cm] {$s_2$};

	\foreach \i in {1,...,3} {
		\node[draw, shape=rectangle, fill, minimum width=0.25, minimum height=0.25, anchor=center] (cn_\i) at ([shift={(0.5,0)}]\i, 2) {};
	}
	
	\foreach \i in {1,...,3} {
			\fill ([shift={(0.5,0)}]\i,2.8) circle[radius=0.135,fill=black!20] coordinate(parityvn_\i);
			\draw[-] (cn_\i) -- (parityvn_\i);
			\draw[<-] ([shift={(0,0.2)}]parityvn_\i) --++ (0, 0.3) node[above] {$L$};
			
	}
	
		\draw[TUMgreen, thick, densely dotted] (cn_1) -- (vn_0);
		\draw[TUMgreen, thick, densely dotted] (cn_1) -- (vn_1);
		\draw[TUMorange, thick, dashed] (cn_1) -- (vn_2);
		\draw[TUMgreen, thick, densely dotted] (cn_1) -- (vn_3);
		\draw (cn_1) -- (vn_4);
		
		\draw[TUMgreen, thick, densely dotted] (cn_2) -- (vn_0);
		\draw[TUMorange, thick, dashed] (cn_2) -- (vn_2);
		\draw (cn_2) -- (vn_4);
		\draw (cn_2) -- (vn_5);

		\draw[TUMgreen, thick, densely dotted] (cn_3) -- (vn_1);
		\draw[TUMgreen, thick, densely dotted] (cn_3) -- (vn_3);
		\draw (cn_3) -- (vn_5);
\end{scope}
\end{tikzpicture} \vspace{-.5cm}
        \caption{Tanner graph after initializing with the message bits.}
        \label{fig:encgraph:graph1}
    \end{subfigure}
    \hfill
    \begin{subfigure}[b]{.3\linewidth}
        \centering
        \usetikzlibrary{arrows}  

\tikzset{%
          insert new path/.style={%
             insert path={%
                  node[pos=0.7,sloped]{\tikz \draw[#1] (-.15pt,0) -- ++(.15 pt,0);}
                  }
             }
         }

\begin{tikzpicture}
\begin{scope}[scale=0.9, a/.style = {insert new path = {-triangle 90}}]
	
	\foreach \i in {0,...,5} {
		\fill (\i,0) circle[radius=0.135,fill=black!20] coordinate(vn_\i);
	}
	\draw[<-] ([shift={(0,-0.2)}]vn_0) --++ (0,-0.3) node[below] {$\infty$} node[below=0.4cm] {$v_1$};
	\draw[<-] ([shift={(0,-0.2)}]vn_1) --++ (0,-0.3) node[below] {$\infty$} node[below=0.4cm] {$v_2$};
 	\draw[<-] ([shift={(0,-0.2)}]vn_2) --++ (0,-0.3) node[below] {$-\infty$} node[below=0.4cm] {$v_3$};
	\draw[<-] ([shift={(0,-0.2)}]vn_3) --++ (0,-0.3) node[below] {$\infty$} node[below=0.4cm] {$v_4$};
	\draw[<-] ([shift={(0,-0.2)}]vn_4) --++ (0,-0.3) node[below] {$0$} node[below=0.4cm] {$s_1$};
	\draw[<-] ([shift={(0,-0.2)}]vn_5) --++ (0,-0.3) node[below] {$0$} node[below=0.4cm] {$s_2$};

	\foreach \i in {1,...,3} {
		\node[draw, shape=rectangle, fill, minimum width=0.25, minimum height=0.25, anchor=center] (cn_\i) at ([shift={(0.5,0)}]\i, 2) {};
	}
	
	\foreach \i in {1,...,3} {
			\fill ([shift={(0.5,0)}]\i,2.8) circle[radius=0.135,fill=black!20] coordinate(parityvn_\i);
			\draw[-] (cn_\i) -- (parityvn_\i);
			\draw[<-] ([shift={(0,0.2)}]parityvn_\i) --++ (0, 0.3) node[above] {$L$};
			
	}
	
		\draw[TUMgreen, thick, densely dotted] (cn_1) -- (vn_0);
		\draw[TUMgreen, thick, densely dotted] (cn_1) -- (vn_1);
		\draw[TUMorange, thick, dashed] (cn_1) -- (vn_2);
		\draw[TUMgreen, thick, densely dotted] (cn_1) -- (vn_3);
		\draw (cn_1) -- (vn_4);
		
		\draw[TUMgreen, thick, densely dotted] (cn_2) -- (vn_0);
		\draw[TUMorange, thick, dashed] (cn_2) -- (vn_2);
		\draw (cn_2) -- (vn_4);
		\draw (cn_2) -- (vn_5);

		\draw[TUMgreen, thick, densely dotted] (cn_3) -- (vn_1);
		\draw[TUMgreen, thick, densely dotted] (cn_3) -- (vn_3);
		\draw (cn_3) -- (vn_5);

   	\draw[-] (cn_3) -- (vn_5)[a] node[pos=0.7, right] {\small $L$};
   	\draw[-] (cn_1) -- (vn_4)[a] node[pos=0.7, left] {\small $-L$};
   	\draw[-] (cn_2) -- (vn_4)[a] node[pos=0.7, right] {\small $0$};
   	\draw[-] (cn_2) -- (vn_5)[a] node[pos=0.7, below] {\small $0$};

\end{scope}
\end{tikzpicture} \vspace{-.5cm}
        \caption{Tanner graph with check-to-variable messages after the first check node update.}
        \label{fig:encgraph:graph2}
    \end{subfigure}

    \vspace{0.3cm}
    \begin{subfigure}[t]{.3\linewidth}
        \centering
        \usetikzlibrary{arrows}  

\tikzset{%
          insert new path/.style={%
             insert path={%
                  node[pos=0.7,sloped]{\tikz \draw[#1,thick] (-.2pt,0) -- ++(.2 pt,0);}
                  }
             }
         }

\begin{tikzpicture}
\begin{scope}[scale=0.9, a/.style = {insert new path = {-triangle 90}}]
	
	\foreach \i in {0,...,5} {
		\fill (\i,0) circle[radius=0.135,fill=black!20] coordinate(vn_\i);
	}
	\draw[<-] ([shift={(0,-0.2)}]vn_0) --++ (0,-0.3) node[below] {$\infty$} node[below=0.4cm] {$v_1$};
	\draw[<-] ([shift={(0,-0.2)}]vn_1) --++ (0,-0.3) node[below] {$\infty$} node[below=0.4cm] {$v_2$};
	\draw[<-] ([shift={(0,-0.2)}]vn_2) --++ (0,-0.3) node[below] {$-\infty$} node[below=0.4cm] {$v_3$};
	\draw[<-] ([shift={(0,-0.2)}]vn_3) --++ (0,-0.3) node[below] {$\infty$} node[below=0.4cm] {$v_4$};
	\draw[<-] ([shift={(0,-0.2)}]vn_4) --++ (0,-0.3) node[below] {$0$} node[below=0.4cm] {$s_1$};
	\draw[<-] ([shift={(0,-0.2)}]vn_5) --++ (0,-0.3) node[below] {$\infty$} node[below=0.4cm] {$s_2$};

	\foreach \i in {1,...,3} {
		\node[draw, shape=rectangle, fill, minimum width=0.25, minimum height=0.25, anchor=center] (cn_\i) at ([shift={(0.5,0)}]\i, 2) {};
	}
	
	\foreach \i in {1,...,3} {
			\fill ([shift={(0.5,0)}]\i,2.8) circle[radius=0.135,fill=black!20] coordinate(parityvn_\i);
	}
    \draw[-] (cn_1) -- (parityvn_1);
    \draw[-] (cn_2) -- (parityvn_2);
	\draw[<-] ([shift={(0,0.2)}]parityvn_1) --++ (0, 0.3) node[above] {$L$};
	\draw[<-] ([shift={(0,0.2)}]parityvn_2) --++ (0, 0.3) node[above] {$L$};
	\draw[->] ([shift={(0,0.2)}]parityvn_3) --++ (0, 0.3) node[above] {$\infty$};
	\draw[TUMgreen, thick, densely dotted] (cn_3) -- (parityvn_3);
	
		\draw[TUMgreen, thick, densely dotted] (cn_1) -- (vn_0);
		\draw[TUMgreen, thick, densely dotted] (cn_1) -- (vn_1);
		\draw[TUMorange, thick, dashed] (cn_1) -- (vn_2);
		\draw[TUMgreen, thick, densely dotted] (cn_1) -- (vn_3);
		\draw (cn_1) -- (vn_4);
		
		\draw[TUMgreen, thick, densely dotted] (cn_2) -- (vn_0);
		\draw[TUMorange, thick, dashed] (cn_2) -- (vn_2);
		\draw (cn_2) -- (vn_4);
		\draw[TUMgreen, thick, densely dotted] (cn_2) -- (vn_5);

		\draw[TUMgreen, thick, densely dotted] (cn_3) -- (vn_1);
		\draw[TUMgreen, thick, densely dotted] (cn_3) -- (vn_3);
		\draw[TUMgreen, thick, densely dotted] (cn_3) -- (vn_5);

\end{scope}
\end{tikzpicture} \vspace{-.5cm}
        \caption{Tanner graph after deciding the first shaping bit.}
        \label{fig:encgraph:graph3}
    \end{subfigure}
    \hfill
    \begin{subfigure}[t]{.3\linewidth}
        \centering
        \usetikzlibrary{arrows}  

\tikzset{%
          insert new path/.style={%
             insert path={%
                  node[pos=0.7,sloped]{\tikz \draw[#1] (-.1pt,0) -- ++(.1 pt,0);}

                  }
             }
         }

\begin{tikzpicture}
\begin{scope}[scale=0.9, a/.style = {insert new path = {-triangle 90}}]
	
	\foreach \i in {0,...,5} {
		\fill (\i,0) circle[radius=0.135,fill=black!20] coordinate(vn_\i);
	}
	\draw[<-] ([shift={(0,-0.2)}]vn_0) --++ (0,-0.3) node[below] {$\infty$} node[below=0.4cm] {$v_1$};
	\draw[<-] ([shift={(0,-0.2)}]vn_1) --++ (0,-0.3) node[below] {$\infty$} node[below=0.4cm] {$v_2$};
	\draw[<-] ([shift={(0,-0.2)}]vn_2) --++ (0,-0.3) node[below] {$-\infty$} node[below=0.4cm] {$v_3$};
	\draw[<-] ([shift={(0,-0.2)}]vn_3) --++ (0,-0.3) node[below] {$\infty$} node[below=0.4cm] {$v_4$};
	\draw[<-] ([shift={(0,-0.2)}]vn_4) --++ (0,-0.3) node[below] {$0$} node[below=0.4cm] {$s_1$};
	\draw[<-] ([shift={(0,-0.2)}]vn_5) --++ (0,-0.3) node[below] {$\infty$} node[below=0.4cm] {$s_2$};

	\foreach \i in {1,...,3} {
		\node[draw, shape=rectangle, fill, minimum width=0.25, minimum height=0.25, anchor=center] (cn_\i) at ([shift={(0.5,0)}]\i, 2) {};
	}
	
	\foreach \i in {1,...,3} {
			\fill ([shift={(0.5,0)}]\i,2.8) circle[radius=0.135,fill=black!20] coordinate(parityvn_\i);
	}
    \draw[-] (cn_1) -- (parityvn_1);
    \draw[-] (cn_2) -- (parityvn_2);
	\draw[<-] ([shift={(0,0.2)}]parityvn_1) --++ (0, 0.3) node[above] {$L$};
	\draw[<-] ([shift={(0,0.2)}]parityvn_2) --++ (0, 0.3) node[above] {$L$};
	\draw[->] ([shift={(0,0.2)}]parityvn_3) --++ (0, 0.3) node[above] {$\infty$};
	\draw[TUMgreen, thick, densely dotted] (cn_3) -- (parityvn_3);
	
		\draw[TUMgreen, thick, densely dotted] (cn_1) -- (vn_0);
		\draw[TUMgreen, thick, densely dotted] (cn_1) -- (vn_1);
		\draw[TUMorange, thick, dashed] (cn_1) -- (vn_2);
		\draw[TUMgreen, thick, densely dotted] (cn_1) -- (vn_3);
		\draw (cn_1) -- (vn_4);
		
		\draw[TUMgreen, thick, densely dotted] (cn_2) -- (vn_0);
		\draw[TUMorange, thick, dashed] (cn_2) -- (vn_2);
		\draw (cn_2) -- (vn_4);
		\draw[TUMgreen, thick, densely dotted] (cn_2) -- (vn_5);

		\draw[TUMgreen, thick, densely dotted] (cn_3) -- (vn_1);
		\draw[TUMgreen, thick, densely dotted] (cn_3) -- (vn_3);
		\draw[TUMgreen, thick, densely dotted] (cn_3) -- (vn_5);

   	\draw[-] (cn_1) -- (vn_4)[a] node[pos=0.7, left] {\small $-L$};
   	\draw[-] (cn_2) -- (vn_4)[a] node[pos=0.7, right] {\small $-L$};

\end{scope}
\end{tikzpicture} \vspace{-.5cm}
        \caption{Tanner graph with check-to-variable messages after the second check node update.}
        \label{fig:encgraph:graph4}
    \end{subfigure}
    \hfill
    \begin{subfigure}[t]{.3\linewidth}
        \centering
        \usetikzlibrary{arrows}  

\tikzset{%
          insert new path/.style={%
             insert path={%
                  node[pos=0.7,sloped]{\tikz \draw[#1,thick] (-.2pt,0) -- ++(.2 pt,0);}
                  }
             }
         }

\begin{tikzpicture}
\begin{scope}[scale=0.9, a/.style = {insert new path = {-triangle 90}}]
	
	\foreach \i in {0,...,5} {
		\fill (\i,0) circle[radius=0.135,fill=black!20] coordinate(vn_\i);
	}
	\draw[<-] ([shift={(0,-0.2)}]vn_0) --++ (0,-0.3) node[below] {$\infty$} node[below=0.4cm] {$v_1$};
	\draw[<-] ([shift={(0,-0.2)}]vn_1) --++ (0,-0.3) node[below] {$\infty$} node[below=0.4cm] {$v_2$};
	\draw[<-] ([shift={(0,-0.2)}]vn_2) --++ (0,-0.3) node[below] {$-\infty$} node[below=0.4cm] {$v_3$};
	\draw[->] ([shift={(0,-0.2)}]vn_3) --++ (0,-0.3) node[below] {$\infty$} node[below=0.4cm] {$v_4$};
	\draw[->] ([shift={(0,-0.2)}]vn_4) --++ (0,-0.3) node[below] {$-\infty$} node[below=0.4cm] {$s_1$};
	\draw[->] ([shift={(0,-0.2)}]vn_5) --++ (0,-0.3) node[below] {$\infty$} node[below=0.4cm] {$s_2$};

	\foreach \i in {1,...,3} {
		\node[draw, shape=rectangle, fill, minimum width=0.25, minimum height=0.25, anchor=center] (cn_\i) at ([shift={(0.5,0)}]\i, 2) {};
	}
	
	\foreach \i in {1,...,3} {
			\fill ([shift={(0.5,0)}]\i,2.8) circle[radius=0.135,fill=black!20] coordinate(parityvn_\i);
			\draw[TUMgreen, thick, densely dotted] (cn_\i) -- (parityvn_\i);
	}
	\draw[->] ([shift={(0,0.2)}]parityvn_1) --++ (0, 0.3) node[above] {$\infty$};
	\draw[->] ([shift={(0,0.2)}]parityvn_2) --++ (0, 0.3) node[above] {$\infty$};
	\draw[->] ([shift={(0,0.2)}]parityvn_3) --++ (0, 0.3) node[above] {$\infty$};

		\draw[TUMgreen, thick, densely dotted] (cn_1) -- (vn_0);
		\draw[TUMgreen, thick, densely dotted] (cn_1) -- (vn_1);
		\draw[TUMorange, thick, dashed] (cn_1) -- (vn_2);
		\draw[TUMgreen, thick, densely dotted] (cn_1) -- (vn_3);
		\draw[TUMorange, thick, dashed] (cn_1) -- (vn_4);
		
		\draw[TUMgreen, thick, densely dotted] (cn_2) -- (vn_0);
		\draw[TUMorange, thick, dashed] (cn_2) -- (vn_2);
		\draw[TUMorange, thick, dashed] (cn_2) -- (vn_4);
		\draw[TUMgreen, thick, densely dotted] (cn_2) -- (vn_5);

		\draw[TUMgreen, thick, densely dotted] (cn_3) -- (vn_1);
		\draw[TUMgreen, thick, densely dotted] (cn_3) -- (vn_3);
		\draw[TUMgreen, thick, densely dotted] (cn_3) -- (vn_5);
\end{scope}
\end{tikzpicture} \vspace{-.5cm}
        \caption{Tanner graph after deciding the second shaping bit.}
        \label{fig:encgraph:graph5}
    \end{subfigure}
    \caption{Example of the encoding procedure with $n_c=9$, $k_c=6$, $\ell = 2$, and DM output $\bm{v} = [0, 0, 1, 0]$. The subfigures show (a) the generator matrix of the code and (b)-(f) the steps of the proposed encoding algorithm.}
    \label{fig:encgraphs}
\end{figure*}

\subsection{Principles of Shaped LDPC Codes}
We propose an efficient \ac{BP}-like algorithm for \ac{LLPS} that approximates the solution of \eqref{eq:LLPSopt}. Consider an $(n_k, k_c)$ \ac{LDPC} code with parity-check matix $\bm{H}$ and systematic generator matrix $\bm{G}_\text{sys} = [\bm{I}_{k_c} \,|\, \bm{G}_\text{p}]$ where $\bm{I}_{k_c}$ is the $k_c \times k_c$ identity matrix and $\bm{G}_\text{p}$ is a $k_c \times (n_c-k_c)$ matrix. The codeword $\bm{c}$ is related to the input $\bm{u}$ by
\begin{equation}
    \bm{c} = \bm{u} \bm{G} = [\bm{u} \,|\, \bm{u}\bm{G}_\text{p}].
\end{equation}

We reserve $\ell$ of the $k_c$ systematic bits for shaping. These shaping bits are a function of the remaining $k_c - \ell$ systematic bits and chosen such that the parity bits $\bm{u}\bm{G}_\text{p}$ have a non-uniform distribution. The encoding procedure to determine the shaping bits is described in the next subsection. %

Since we determine bits from the systematic part of the code, any choice of the shaping bits gives a valid codeword. The remaining $k_c - \ell$ systematic bits are assumed to be the output of a \ac{DM} (we use the \ac{CCDM} \cite{CCDM_schulte}) with rate $R_\text{DM} = k / (k_c - \ell)$, where $k$ is the length of the uniform information sequence to be transmitted. The overall transmission rate is
\begin{align*}
    R = k / n_c  = R_\text{DM} (k_c - \ell) / n_c = R_\text{DM} R_\text{c} - \ell/n_c .
\end{align*}
where $R_c = k_c/n_c$ is the code rate of the LDPC code.

\subsection{Encoding}
The algorithm encodes on the Tanner graph of the generator matrix using a BP-like algorithm that successively determines the $\ell$ shaping bits using $\ell$ iterations over the graph, i.e., each iteration determines one shaping bit. Fig.~\ref{fig:encgraphs} shows an example with $n_c=9$, $k_c=6$, and $\ell = 2$.

The \acp{VN} of the Tanner graph are divided into three groups:
\begin{itemize}
    \item \emph{Message VNs:} the $k_c - \ell$ systematic \acp{VN} of the \ac{DM} output $\bm{v}$ are fixed by setting the \ac{LLR} $L_i$ of \ac{VN} $i$ to $+\infty$ or $-\infty$ for a DM output $0$ or $1$, respectively.
    \item \emph{Shaping VNs:} the $\ell$ systematic \acp{VN} of the shaping bits are initialized with the \ac{LLR} value $L_i=0$.
    \item \emph{Parity VNs:} the $n_c - k_c$ \acp{VN} of the parity bits have \ac{LLR} $L_i = \log(p_0/(1-p_0)) \triangleq L$, which corresponds to the target distribution.
\end{itemize}
Fig.~\ref{fig:encgraph:graph1} depicts the systematic message and shaping \acp{VN} on the bottom and the parity VNs on the top.

In the example, the \acp{LLR} of the message \acp{VN} are initialized according to the \ac{DM} output $\bm{v} = [0, 0, 1, 0]$. %
All check-to-variable node messages are initialized with $0$ and all variable-to-check node messages are initialized with the \ac{LLR} of the variable node. Thus, all messages from the message \acp{VN} are $\pm \infty$, all messages from the shaping \acp{VN} are $0$, and all messages from the parity \acp{VN} are $L$.
Edges that carry messages from fixed \acp{VN} are marked as green-dotted ($L_i=+\infty$) or red-dashed ($L_i=-\infty$) in Fig.~\ref{fig:encgraphs}. These edges do not need to be updated and have a limited impact in the check node operations.

The algorithm now performs $\ell$ iterations on the graph to determine the $\ell$ shaping bits. Each iteration consists of the following steps.
\begin{itemize}
    \item \emph{CN updates:} update the check-to-variable messages to via \eqref{eq:cnupdate}. The update rule has three possible outcomes: if a CN has more than one edges to undetermined shaping VNs, all messages are $0$. If a CN has exactly one edge to an undetermined shaping VN, then the message on this edge will be $+L$ or $-L$, depending on whether the number of incoming $-\infty$-message is even or odd; see~Fig.~\ref{fig:encgraph:graph2}.
    \item \emph{Determine one shaping bit (decimation step):} for all undetermined shaping VNs $i$, calculate
    \begin{equation}
        \widetilde{L}_i^{\text{APP}} = L_i^\text{APP} + L \label{eq:Ls}
    \end{equation} 
    and choose one of the VNs with largest $|\widetilde{L}_i^{\text{APP}}|$. Fix the corresponding shaping bit using
    \begin{equation}
        s_i = \begin{cases}
            0, \quad  \widetilde{L}_i^{\text{APP}} \geq 0 \\
            1, \quad \text{else}
        \end{cases}
    \end{equation}
    and change the LLR $L_i$ of VN $i$ accordingly. Further, update the messages on the edges connected to this VN. We remark that the $ \widetilde{L}_i^{\text{APP}}$ are always integer multiples of $L$ and can be interpreted as how many parity bits are (in this iteration) set to zero compared to how many are set to one. The offset $+L$ in $ \widetilde{L}_i^{\text{APP}}$ is applied in order to induce a bias on the distribution of the shaping bits as well.

    In the example, we have $ \widetilde{L}_1^{\text{APP}} = 0$ and $ \widetilde{L}_2^{\text{APP}} = 2L$ and thus fix $s_2 = 0$. %
    The updated graph is shown in Fig.~\ref{fig:encgraph:graph3}. Observe that fixing $s_2=0$ determines the third parity bit to be $0$, as illustrated by the outgoing APP LLR of $+\infty$. %
\end{itemize}
After $\ell$ iterations, the algorithm determined all $\ell$ shaping bits, and in a final stage it calculates all parity bits and puts out a valid codeword with a non-uniform distribution.

\subsection{Decoding}
Since all shaping bits are in the systematic part of the codeword
we need not modify the decoder. After decoding, one simply discards the shaping bits. We remark that the shaping bits could be used for error detection by re-encoding the message and comparing the decoded shaping bits with the re-encoded ones. %

\subsection{Discussion}

The algorithm is based on the Tanner graph of the systematic generator matrix instead of the parity check matrix. The former matrix is usually denser than the latter. From the perspective of computational complexity this is not an issue as the CN update \eqref{eq:cnupdate} only needs to be calculated if the effective CN degree is $1$. However, if all check-to-variable node messages are $0$, then the algorithm must guess one bit. The algorithm might thus perform poorly if the generator matrix is excessively dense or if one approximates an extreme distribution with many shaping bits. The behaviour improves by optimizing the decimation strategy and the positions of the shaping bits, by designing tailored codes, or by using more complex decimation strategies. For example, one may choose multiple shaping bits jointly or use brute-force search for the first bits and encode the remaining bits for multiple choices.

\section{Simulation Results} \label{sec:results}
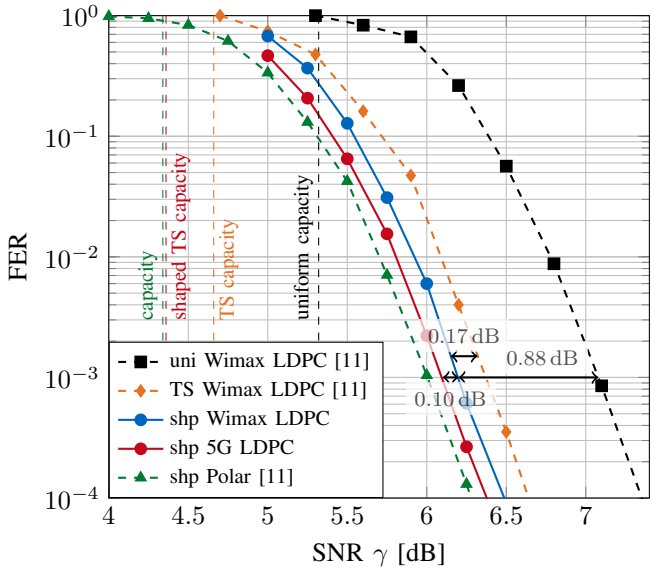
\begin{figure}[t]
    \centering
    \pgfplotsset{compat=newest}
\pgfplotsset{set layers}%

\begin{tikzpicture} %
    \begin{axis}
        [	height=8cm,
        	width=.99\linewidth,
            grid=both,
            ymode=log,
            xlabel={SNR $\gamma$ [dB]},
            ylabel={FER},
            ymax=1,
            ymin=1e-4,
             xmin=4,
             xmax=7.4,
            legend cell align={left},
            legend style={at={(0,0)},anchor=south west,font=\footnotesize,} %
        ]
        \addplot[thick, black, dashed, mark=square*,mark options={solid}] table [x=snr, y=fer] {res/wimax-uni.txt};
        \label{curve:23:uniLDPC}
        \addlegendentry{uni Wimax LDPC \cite{wiegart_shaped19}};

        \addplot[thick, TUMorange, dashed, mark=diamond*, ,mark options={solid}] table [x=snr, y=fer] {res/wimax-ts1.txt};
        \label{curve:23:TSLDPC}
	    \addlegendentry{\footnotesize TS Wimax LDPC \cite{wiegart_shaped19}};
        
        \addplot[thick, TUMblue, mark=*] table [x=SNR, y=FER] {res/wimax-shp-1056_34B_s10.txt};
        \label{curve:23:shpLDPC}
	       \addlegendentry{shp Wimax LDPC} %

        \addplot[thick, TUMred, mark=*] table [x=SNR, y=FER] {res/5G/shp-5G_BG1_n1008_0.78_s1-32.txt};
        \label{curve:23:5GshpLDPC}
	       \addlegendentry{shp 5G LDPC} %

        \addplot[thick, TUMgreen, mark=triangle*, dashed,mark options={solid}] table [x=SNR, y=FER] {res/polar_1024_683.txt};
        \label{curve:23:shpPolar}
        \addlegendentry{shp Polar \cite{wiegart_shaped19} } %

        \begin{pgfonlayer}{axis foreground}
	      \draw[<->, thick] (6.2,1e-3) -- (7.08, 1e-3) node[pos=0.6, above=0.03cm, fill=white, fill opacity=0.7, text opacity=1] {\footnotesize $\SI{0.88}{dB}$};
       \draw[<->, thick] (6.15,1.5e-3) -- (6.32, 1.5e-3) node[midway, above=0.05cm, fill=white, fill opacity=0.7, text opacity=1] {\footnotesize $\SI{0.17}{dB}$};
       \draw[<->, thick] (6.1,1e-3) -- (6.2, 1e-3) node[midway, below=0.05cm, fill=white, fill opacity=0.7, text opacity=1] {\footnotesize $\SI{0.10}{dB}$};
	      \end{pgfonlayer}

  \draw[dashed, black] (5.32, 1) -- (5.32, 1e-4) node[pos=0.65, rotate=90, right, yshift=0.2cm] {\footnotesize uniform capacity};
  \draw[dashed, TUMgreen] (4.34, 1) -- (4.34, 1e-4) node[pos=0.65, rotate=90, right, yshift=0.2cm] {\footnotesize capacity}; %
  \draw[dashed, TUMred] (4.36, 1) -- (4.36, 1e-4) node[pos=0.65, rotate=90, right, yshift=-0.2cm] {\footnotesize shaped TS capacity}; %
  \draw[dashed, TUMorange] (4.66, 1) -- (4.66, 1e-4) node[pos=0.65, rotate=90, right, yshift=-0.2cm] {\footnotesize TS capacity}; %
  
    \end{axis}
\end{tikzpicture}%
    \caption{FER with overall rate $R=2/3$ and $n_c\approx\SI{1000}{bits}$. The uniform (\ref{curve:23:uniLDPC}) and TS (\ref{curve:23:TSLDPC}) Wimax LDPC codes have $n_c=1056$ and are taken from \cite{wiegart_shaped19}. The TS scheme has code rate $R_c = 0.75$. The shaped Wimax LDPC code (\ref{curve:23:shpLDPC}) has $n_c=1056$, $R_c = 0.75$ and $\ell = 10$ shaping bits. The shaped 5G LDPC code (\ref{curve:23:5GshpLDPC}) has $n_c=1008$, $R_c = 0.78$, and $\ell=32$. The shaped polar code (\ref{curve:23:shpPolar}) is from \cite{wiegart_shaped19} and has $n_c=1024$.}
    \label{fig:sim:R23}
\end{figure}

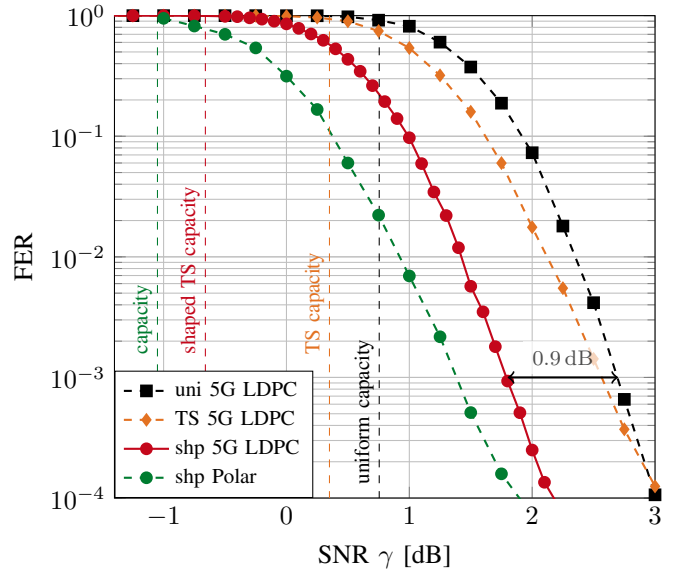
\begin{figure}[t]
    \centering
    \pgfplotsset{compat=newest}
\pgfplotsset{set layers}%

\begin{tikzpicture} %
    \begin{axis}
        [	height=8cm,
        	width=.99\linewidth,
            grid=both,
            ymode=log,
            xlabel={SNR $\gamma$ [dB]},
            ylabel={FER},
            ymax=1,
            ymin=1e-4,
             xmin=-1.4,
             xmax=3.0,
            legend cell align={left},
            legend style={at={(0,0)},anchor=south west,font=\footnotesize,} %
        ]
    \addplot[thick, black, dashed, mark=square*, ,mark options={solid}] table [x=SNR, y=FER] {res/5G/uni-5G_BG1_n1056_R13.txt};
            \label{curve:13:uniLDPC}
	    \addlegendentry{\footnotesize uni 5G LDPC}; %

    \addplot[thick, TUMorange, dashed, mark=diamond*,mark options={solid}] table [x=SNR, y=FER] {res/5G/ts1-5G_BG1_n1056_R13.txt};
    \label{curve:13:TSLDPC}
	    \addlegendentry{TS 5G LDPC}; %
     
     \addplot[thick, TUMred, mark=*] table [x=SNR, y=FER] {res/5G/shp-5G_BG1_n1056_R13_s1-64.txt};
     \label{curve:13:5GshpLDPC}
	    \addlegendentry{shp 5G LDPC}; %

     \addplot[thick, TUMgreen, dashed, mark=*, mark options={solid}] table [x=SNR, y=FER] {res/polar_1024_342.txt};
     \label{curve:13:shpPolar}
	    \addlegendentry{shp Polar} %

    \draw[dashed, TUMgreen] (-1.05, 1) -- (-1.05, 1e-4) node[pos=0.72, rotate=90, right, yshift=0.2cm] {\footnotesize capacity};

    \draw[dashed, TUMred] (-0.66, 1) -- (-0.66, 1e-4) node[pos=0.72, rotate=90, right, yshift=0.2cm] {\footnotesize shaped TS capacity};

    \draw[dashed, TUMorange] (0.35, 1) -- (0.35, 1e-4) node[pos=0.72, rotate=90, right, yshift=0.2cm] {\footnotesize TS capacity};

    \draw[dashed, black] (0.755, 1) -- (0.755, 1e-4) node[pos=0.99, rotate=90, right, yshift=0.2cm] {\footnotesize uniform capacity};

    \begin{pgfonlayer}{axis foreground}
	      \draw[<->, thick] (1.8,1e-3) -- (2.7, 1e-3) node[pos=0.5, above=0.03cm, fill=white, fill opacity=0.7, text opacity=1] {\footnotesize $\SI{0.9}{dB}$};
	\end{pgfonlayer}

    \end{axis}
\end{tikzpicture}%
    \caption{FER with overall rate $R=1/3$ and $n_c\approx\SI{1000}{bits}$. The LDPC codes are from the 5G standard and have $n_c=1056$. The TS LDPC code (\ref{curve:13:TSLDPC}) has $R_c = 1/2$ (according to \cite[Table I]{git_protograph19}). The shaped LDPC code (\ref{curve:13:5GshpLDPC}) has $R_c = 2/3$, $64$ punctured bits, and $\ell=64$ shaping bits. 
    The shaped polar code (\ref{curve:13:shpPolar}) has $n_c=1024$ and uses both SCL encoding and decoding with $L=32$.}
    \label{fig:sim:R13}
\end{figure}

We simulated performance over an AWGN channel with OOK modulation, as described in Sec.~\ref{sec:awgn}. We compare to the polar codes from \cite{wiegart_shaped19} and the \ac{TS} scheme from \cite{git_protograph19} with LDPC codes.

For some curves we used 5G LDPC codes \cite{5g}. They are all obtained from base graph $1$. According to the standard, some of the systematic bits are punctured. We place the shaping bits within the punctured bits and slightly modify \eqref{eq:Ls} to $ \widetilde{L}_i^{\text{APP}} = L_i^\text{APP}$, i.e., we discard the additional offset of $+L$ as the punctured bits do not need to have a non-uniform distribution. If there are more punctured bits than shaping bits, the remaining punctured bits are filled with information bits.

Fig.~\ref{fig:sim:R23} considers the transmission rate $R=2/3$ and compares the \acp{FER} of the proposed algorithm to uniform signaling, the \ac{TS} scheme from \cite{git_protograph19}, and shaped polar codes. With LDPC codes from the Wimax standard \cite{ieee_wimax} we gain approximately $\SI{0.88}{dB}$ compared to uniform signaling and $\SI{0.17}{dB}$ compared to the \ac{TS} scheme. For the shaped and TS curves we used $R_c = 0.75$; the shaped curve was generated using $\ell=10$ shaping bits. We can further improve by using a 5G LDPC code. We chose a code with $n_c=1008$ ($72$ punctured bits) and used $\ell=32$. This gives an additional $\SI{0.1}{dB}$ of gain compared to the shaped Wimax LDPC code. We further depict the shaped polar code from \cite[Fig.~5]{wiegart_shaped19} for reference. The polar code has $n_c=1024$, is combined with an outer $\SI{16}{bit}$ CRC, and a \ac{SCL} decoder \cite{tal2015_list} with list size $L=32$ was used for encoding and for decoding. Both schemes perform very close to each other, but the polar code performs slightly better. This is expected for such rather short block lengths and we remark that the polar code has a tailored and optimized design, while the off-the-shelf LDPC code was not designed for this purpose.

The figure also depicts achievable rates as vertical dashed lines. As our scheme also works in a \ac{TS} manner (depending on $\ell$, the distribution of the systematic part and the distribution of the parity part are not necessarily the same), we depict the achievable rate with the empirical distributions created by our algorithm as ``shaped TS capacity''. For this scenario, the achievable rate of our scheme is very close to capacity and the gains observed in the \ac{FER} curves match the theoretical predictions. 

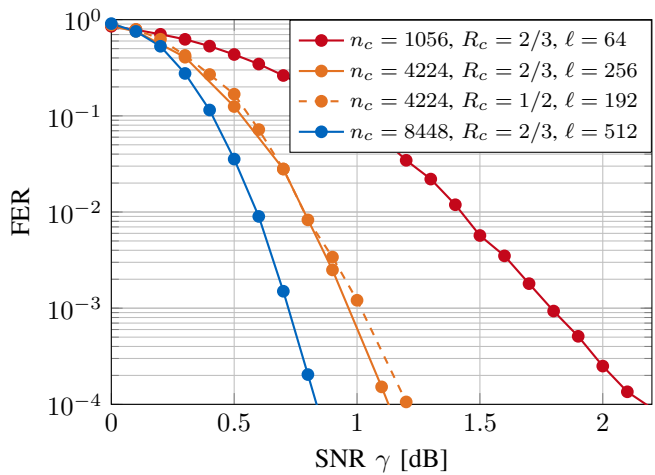
\begin{figure}[t]
    \centering
    \pgfplotsset{compat=newest}
\pgfplotsset{set layers}%

\begin{tikzpicture} %
    \begin{axis}
        [	height=6.7cm,
        	width=.99\linewidth,
            grid=both,
            ymode=log,
            xlabel={SNR $\gamma$ [dB]},
            ylabel={FER},
            ymax=1,
            ymin=1e-4,
             xmin=0,
             xmax=2.2,
            legend cell align={left},
            legend style={at={(1,1)},anchor=north east,font=\footnotesize,} %
        ]

     \addplot[thick, TUMred, mark=*] table [x=SNR, y=FER] {res/5G/shp-5G_BG1_n1056_R13_s1-64.txt};
     \label{curve:13:5GshpLDPC}
	    \addlegendentry{$n_c=1056$, $R_c=2/3$, $\ell=64$}; %

    \addplot[thick, TUMorange, mark=*] table [x=SNR, y=FER] {res/5G/shp-5G_BG1_n4224_R13_s1-256.txt};
	    \addlegendentry{$n_c=4224$, $R_c = 2/3$, $\ell=256$};

     \addplot[thick, TUMorange, dashed, mark=*, mark options={solid}] table [x=SNR, y=FER] {res/5G/shp-5G_BG1_n4224_R13_Rc12_s1-192.txt};
	    \addlegendentry{$n_c=4224$, $R_c = 1/2$, $\ell=192$};

    \addplot[thick, TUMblue, mark=*] table [x=SNR, y=FER] {res/5G/shp-5G_BG1_n8448_R13_Rc23_s1-512.txt};
	    \addlegendentry{$n_c=8448$, $R_c = 2/3$, $\ell=512$};

    \draw[dashed, TUMgreen] (-1.05, 1) -- (-1.05, 1e-4) node[pos=0.65, rotate=90, right, yshift=0.2cm] {\footnotesize capacity};

    \draw[dashed, TUMred] (-0.66, 1) -- (-0.66, 1e-4) node[pos=0.65, rotate=90, right, yshift=0.2cm] {\footnotesize shaped TS capacity};

    \begin{pgfonlayer}{axis foreground}
	\end{pgfonlayer}

    \end{axis}
\end{tikzpicture}
    \caption{FER with overall rate $R=1/3$ of shaped 5G LDPC codes with different code lengths and code rates.}
    \vspace{-0.2cm}
    \label{fig:sim:R13differentlength}
\end{figure}
Fig.~\ref{fig:sim:R13} shows results for a transmission rate of $R=1/3$. Here, the theoretical shaping gain is $\SI{1.8}{dB}$. The shaping gain of the TS scheme is small due to the high number of uniform parity bits for low transmission rates \cite{git_protograph19}. 
The proposed scheme with a 5G code ($64$ punctured bits and $\ell=64$) gains $\SI{0.9}{dB}$ compared to uniform transmission. This is a bit less then it could be expected from the achievable rates (and we also observe that the FER waterfall is not accurately predicted by the capacity for the shaped curve). We believe that this is a code design issue and further gains are possible by designing tailored codes. We remark that the target distribution for this scenario is quite biased ($p_0=0.83$) and our algorithm is not fully capable of creating this distribution. Thus there is a gap to capacity of approximately $\SI{0.4}{dB}$ (see "shaped TS capacity" in Fig.~\ref{fig:sim:R13}).
The polar-coded curve was designed as in \cite{wiegart_shaped19} and we used an $\SI{8}{bit}$ CRC, $|\cD| = 242$, and \ac{SCL} encoding and decoding with $L=32$. The polar code outperforms the proposed LDPC code (the worse slope could potentially be improved by optimizing the CRC or using a larger list size).

Finally, Fig.~\ref{fig:sim:R13differentlength} shows the FER performance of our encoding scheme with $R=1/3$ for codes of different length. We see that the proposed scheme works for different lengths (we depict $n_c=1056$, $n_c=4224$, and $n_c=8448$). The fraction of shaping bits compared to the code length was kept constant and for all three cases we obtain the same empirical distribution of the parity bits. For $n_c=4224$ we further depict a curve for a code with a lower code rate. The obtained distribution is naturally less biased but the code has better error correcting capability and the two curves for $n_c=4224$ lie very close to each other. This shows that our scheme can, similarly to \ac{PAS}, also be used for rate adaptivity.

\section{Conclusion} \label{sec:conclusion}
We proposed an efficient BP-like algorithm for LLPS using LDPC codes based on the Tanner graph of the systematic generator matrix. It is sub-optimal, but still delivers a good FER performance in simulations with standard LDPC codes. The algorithm allows for rate flexibility and compared to other schemes for LDPC codes it always generates a valid codeword.
Further work to optimize the algorithm can include tailored code-design and optimizing the decimation strategy and the selection of the shaping bits.

\section*{Acknowledgment}
The authors thank Gerhard Kramer for helpful comments. %

\bibliographystyle{IEEEtran}
\bibliography{Linfang_TCOM,literature}

\end{document}